\documentclass[journal=jpcbfk,manuscript=article]{achemso}

\usepackage{graphicx}
\graphicspath{ {./images/} }
\usepackage{gensymb}
\usepackage{xcolor}
\usepackage{ulem}
\usepackage{comment}
\usepackage{float}
\usepackage{xr}
\usepackage{hyperref}

\usepackage{amsmath}
\newcommand{\fourplus}[1]{#1\ensuremath{^{4+}}\kern-\scriptspace}

\title[An \textsf{achemso} demo]{Opportunities and Challenges in Unsupervised Learning: The Case of Aqueous Electrolyte Solutions}

\author{Giulia Sormani}
\affiliation{The ``Abdus Salam'' International Centre for Theoretical Physics, I-34151 Trieste, Italy}

\author{Alex Rodriguez}
\email{alexdepremia@gmail.com}
\affiliation{University of Trieste, Dipartimento di Matematica e Geoscienze, I-34151 Trieste, Italy.}

\author{Ali Hassanali}
\email{ahassana@ictp.it}
\affiliation{The ``Abdus Salam'' International Centre for Theoretical Physics, I-34151 Trieste, Italy}

\begin{document}

\begin{abstract}
Machine learning has emerged as a powerful tool in atomistic simulations, enabling the identification of complex patterns in molecular systems limiting human intervention and bias. However, the practical implementation of these methods presents significant technical challenges, particularly in the selection of hyperparameters and in the physical interpretability of machine-learned descriptors. In this work, we systematically investigate these challenges by applying an unsupervised learning protocol to a fundamental problem in physical chemistry namely, how ions perturb the local structure of water. Using the Smooth Overlap of Atomic Positions(SOAP) descriptors, we demonstrate how the intrinsic dimension (ID) serves as a guide for selecting hyperparameters and interpreting structural complexity.
Furthermore, we construct a high-dimensional free energy landscape encompassing all water environments surrounding different ions. This analysis reveals how the physical properties of ions are intricately reflected in their hydration shells, shaping the landscape through specific connections between different minima.
Our findings highlight the difficulty in balancing algorithmic automation with the need of employing both physical and chemical intuition, particularly for the construction of meaningful descriptors and for the interpretation of final results. By critically assessing the methodological hurdles associated with unsupervised learning, we provide a road map for researchers looking to harness these techniques for studying electrolyte and aqueous solutions in general. 
\end{abstract}

\section{Introduction}

Machine learning has revolutionized the physical sciences over the past decade\cite{2023_ML_molecular_sim}. In the field of atomistic simulations, this has led to two major developments. Firstly, the construction of ML-based interatomic potentials which has enabled simulations with the accuracy comparable to electronic structure methods such as density functional theory (DFT), while at the same time, significantly extending the timescales accessible beyond tens of picoseconds\cite{ML_potential_2024,ML_potential_practical_guide_2025}. The second aspect is rooted in the rise of both supervised and unsupervised learning methods applied to data from molecular simulations in order to yield new insights into the complexity of chemical and biological systems. Several studies have demonstrated the effectiveness of these approaches\cite{collective_motions_in_proteins_first_PCA_1991, essential_dynamics_of_proteins_PCA_1993, FES_PCA_stock_2005, isomap_2010, kernel_PCA_2019_bingqing,kernel_PCA_2019_ceriotti, DBscan_application_RNA_2014, DBscan_application_binding_sites_2013, Pande_MSM_2013,MSM_2014, kmeans_2017_noe, protein_association_noe_2017, vampnets_noe_2018, pinamonti_2019_RNA_fraying, kpeaks_2019, carli_sarscov2_2020, TICA_2011,TICA_2014, vampnets_2018, pavan_2023_metal_surface, capelli2022_ice_environments, Pande_2014, ML_ubiquitin_2020, Bingqing_ice_phases_2020, banerjee2024}.
The novelty of these techniques lies in their ability to uncover patterns in the data automatically rather than imposing the users chemical and physical bias a priori, and at the same time, providing a more rigorous statistical framework to quantify the importance and relevance of the underlying patterns in molecular data.

Liquids provide particularly interesting and challenging physical conditions in which to discover patterns. More specifically, due to the large fluctuations and disorder, being able to quantify and probe underlying structures on different length and time-scales has been the subject of numerous studies and lively debate in the field\cite{foffi_sciortino_fluctuations_2022,coslovich_2014_correlation_supercooled_liquids, coslovich_2011_correlations_viscous_liquids,dellago2025}.  Over the last few years, our group has developed an unsupervised learning protocol to study different aqueous systems including bulk water at room temperature\cite{Adu_liquid_water2022} and under supercooled conditions\cite{eddy_supercooled_2024}, as well as aqueous solutions of HCl\cite{solana_zundeig}. In summary, the framework we developed involves constructing high-dimensional local atomic descriptors, extracting the intrinsic dimension (ID) and finally, determining free energies in the high-dimensional embedding manifolds. Although this approach is nominally unsupervised, each step involves the selection of hyperparameters which has to be combined with physical and chemical intuition in order to interpret the results.

In this work, we take a deeper dive into the use of unsupervised learning to study a textbook problem in physical chemistry namely, how ions perturb the local structure of water in close vicinity to them. The structural, dynamical and spectroscopic properties of aqueous electrolyte solutions both in the bulk and at interfaces has often been rationalized in terms of the Hofmeister series\cite{original_hofmeister1888,revisiting_original_Hofmeister_2004} - a ranking of ions based on their tendency to salt-in or out proteins from solution\cite{Cremer_hofmeister_review_2006,Hofmeister_review_2022}. In this context, different ions have often been described as \emph{structure makers} or \emph{structure breakers}, a descriptive label that has been used to rationalize empirical measurements coming from experiments examining how different thermodynamic\cite{anions_effect_on_cloud_point_T_of_lysosyme_Cremer2009,surface_tensions_hofmeister_2010, surface_tension_solutions_1968, krestov_thermodynamics_1990,activity_coefficeints_1952,samoilov_1957}, and dynamical\cite{D_over_Dw_experiments_simulations_kim2012,D_over_Dw_experiments_muller1996, B_coef_Marcus_1995,Bcoef_1957,Bcoef_1959} properties change as a function of salt concentration. 

Our results provide a general framework for how one can navigate the challenges in selecting and physically rationalizing the choice of hyperparameters during the unsupervised learning protocol. In particular, we show that the ID presents a rather compelling way to interrogate the complexity and information encoded by the SOAP features\cite{SOAP_2013} for the water around different ions. We show that the appropriate choice of parameters emerges from an iterative step involving the search and identification of physically interpretable quantities such as the water orientational structure, as well as experimental measurements like the Jones Dole B coefficient, both of which we find are linked to the ID. Finally, we construct the high-dimensional free energy landscape in SOAP space of the water environments around the ions revealing the complex topography and pathways that connect the different species.

The paper is organized as follows. The Methods section provides a comprehensive description of the simulation protocol, followed by an overview of the theoretical framework underlying the SOAP descriptors. This section concludes with a detailed explanation of the unsupervised learning procedure, including the ID estimation, determination of the point-free energies, and the clustering approach. The Results section begins with the selection of hyperparameters for constructing SOAP descriptors, guided by the ID ranking of the high-dimensional SOAP feature spaces of the different ions. This is followed by an analysis of the chemical interpretation of these rankings and culminates in the construction and analysis of high-dimensional free energy landscapes of water environments. Final remarks and relevance of the analysis are presented in the Discussion and Conclusion section.

\section{Computational Methods} 
\label{sec:methods}

\subsection{Molecular Dynamics Protocol}

The focus of this study is to characterize the solvation patterns around a series of cations (calcium, magnesium, lithium, potassium, rubidium, cesium) and anions (chloride, fluoride, bromide, and iodide) in water. To achieve this objective, we conduct molecular dynamics simulations of a single ion in a box of water and compare how the ion perturbs the local solvation structure compared to neat bulk water using the GROMACS software\cite{gromacs_2005}. In all our simulations, we use the Madrid-2019 scaled-charges force field\cite{Madrid2019,madrid2019_extension} for ions, combined with the TIP4P-2005 water model\cite{tip4p_2005}. Madrid-2019 is a non-polarizable force field in which all the ionic charges are scaled by a factor of 0.85 which serves as an electronic continuum correction which accounts for electronic polarization in a mean-field way (see ref.  \citenum{Pavel_charge_scaling_manifesto} for a detailed explanation). This scaling results in effective charges of 0.85 and 1.7 (in electron units) for monovalent and divalent ions, respectively.
The charge-scaling approach was originally proposed by Leontyev and Stuchebrukhov\cite{scaling_leontyev2010,scaling_leontyev2011} and has since been widely adopted (see Ref.\citenum{review_biological_applications_charge_scaling} for a comprehensive review) since it overcomes some of the pitfalls of previous standard non-polarizable force fields such as unphysical ion pairing and salt precipitation well below the solubility limit.

A cubic water box of 3x3x3 nm (containing 909 water molecules) is created and a randomly selected water molecule is replaced with the ion of interest. This box size is sufficiently large to avoid significant finite-size effects: as shown in the lower panels of Figure 1, the ion–water hydrogen radial distribution functions converge to unity at a distance smaller than 1 nanometer from the ion, indicating that the correlation length is well below half the box size.
For all the systems, we conduct an equilibration procedure before moving to production runs. The procedure begins with energy minimization followed by a 20ps NVT simulation in which the temperature is gradually brought from 0K to 300K. Finally, the density is equilibrated through a simulation in the NPT ensemble for 2ns using the Parrinello-Rahman barostat\cite{parrinello_Rahman_barostat} fixed at 1 bar. The equilibration phase is followed by a 1$\mu s$ simulation in the NVT ensemble at a temperature of 300K. In all the MD simulations, the time step is set to 2fs and the temperature is controlled using the velocity-rescale thermostat\cite{bussi_thermostat}, using a time constant of 0.1ps . The cutoff for the non-bonded interactions is set to 11$\r{A}$. Particle Mesh Ewald(PME)\cite{PME_electrostatic_1993} is used to treat the long-range part of the Coulomb interactions.

\subsection{Smooth Overlap of Atomic Positions (SOAP)}

We use the Smooth Overlap of Atomic Positions\cite{SOAP_2013} (SOAP) as the local atomic descriptors to describe the chemical environments around the ions.  The core principle behind the SOAP descriptors is to represent the chemical environment surrounding a selected atom as an array of coefficients derived from the atomic density function, $\rho(\overrightarrow{r})$, at the atom's position. This atomic density is constructed as a sum of Gaussian functions, each with a width $\sigma$, centered on the positions of neighboring atoms $i$ within a cutoff radius $R_{cut}$ from the central atom:

\begin{equation}
    \rho^{Z_{i}}(\overrightarrow{r}) = \sum_{i}exp(\frac{-|\overrightarrow{r}-\overrightarrow{r_{i}}|^{2}}{2\sigma^{2}})  
\label{sum_of_gaussians}
\end{equation}

It is worth noting that in a system with multiple atomic species, a specific density is calculated restricted to the atoms belonging to each species. For this reason, the density ($\rho^{Z_{i}}$) is labeled by the species atomic number $Z_{i}$. The atomic density $\rho^{Z_{i}}$ can be expanded in both a radial basis set $g_{n}(r)$ and spherical harmonics $Y_{lm}(\theta,\psi)$ where $c^{Z_{i}}_{nlm}$ are the coefficients of the expansion:
\begin{equation}
    \rho^{Z_{i}}(\overrightarrow{r}) = \sum^{N_{max}}_{n=0}\sum^{L_{max}}_{l=0}\sum^{l}_{m=-l} c^{Z_{i}}_{nlm} g_{n}(r)Y_{lm}(\theta,\psi) 
\label{radial_spherical_harmonics}
\end{equation}

These coefficients are, by construction, invariant in front of the permutation of identical atoms. However, to reach the objective of having a representation invariant in front of rotations and translations of the space, a final transformation is needed:
\begin{equation}
p^{Z_{j}Z{k}}_{nn'l}= \pi \sqrt{\frac{8}{2l+1}}\sum_{m}c^{Z_{j}}_{nlm}c^{Z_{k}}_{n'lm}
\label{final_transf}
\end{equation}
being the so-called power spectrum $p^{Z_{j}Z{k}}_{nn'l}$ the final outcome of all the procedure.

The SOAP mathematical structure thus relies on several key physical considerations, including the chemical species of interest, the spatial extent of the region to be explored, the spatial resolution, and finally, the level of detail in the basis set expansion. These considerations are encoded into a set of tunable parameters, which are detailed as follows:

\begin{itemize}
    \item $R_{cut}$: The cutoff radius defining the neighborhood around the central atom within which neighboring atoms contribute to the local density. 
    \item $\sigma$: The width of the Gaussian function assigned to each atom - smaller values of $\sigma$ correspond to resolving finer spatial resolution.
    \item $N_{max}$ and $L_{max}$:  Parameters that control the extent of the expansion of the radial basis set and the spherical harmonics, respectively.
\end{itemize}

Once these parameters are fixed, the chemical environment around an atom throughout the simulation is represented as a matrix. In this matrix, the rows correspond to the simulation frames sampled at fixed time intervals, and the columns correspond to the components of the SOAP spectrum. The number of components of the SOAP power spectrum is determined by the number of atomic species and the values of $N_{max}$ and $L_{max}$, often reaching the order of thousands.

For this work, the Dscribe package\cite{dscribe2020} is used to compute the SOAP coordinates from MD trajectories of single ions in water and of bulk water. The radial basis set used for SOAP are primitive gaussian type orbitals (GTO). Frames are selected at a time step of 100=ps, for a total of 10K frames over the trajectories of length 1$\mu$s. The selection of the values of all SOAP parameters and how they affect properties of interest will be addressed in the Results section.

\subsection{Intrinsic Dimension Estimator}

Datasets generated from molecular dynamics simulations are typically high-dimensional. In most molecular systems however, correlations between the data points typically lower the dimensionality of the manifold in which the data reside\cite{low_dimensionality_biological_manifold}. The number of independent degrees of freedom needed the describe this lower dimensional manifold is referred to as the Intrinsic Dimension (ID).
In this study, we calculate ID of the SOAP dataset of each ionic system and of bulk water by means of the two-NN estimator\cite{twoNN}. The mathematical formulation of this technique relies on the analysis of the statistical properties of the quantity $\mu_{i}=r_{i,2}/r_{i,1}$ which corresponds, for each point $i$, to the ratio of second nearest neighbor distance and first nearest neighbor distance.  Error bars are statistically derived from a sub-sampling approach:  for each dataset, the ID is computed twice using two randomly selected subsets, each containing half of the original frames, the uncertainty is obtained as the standard deviation of the two obtained IDs. The same procedure can be repeated with a higher number of subsets to verify the ID and uncertainty convergence. A key advantage of the two-NN estimator is that it only requires the density to be constant within the second neighbor's range of each point, making it less sensitive to density inhomogeneities and possible curvature of the data manifold. For the case of the SOAP coordinates, the distance between points is calculated as the Euclidean distance of the two normalized SOAP power spectra. Being that the SOAP arrays are normalized, this distance effectively mimics the Euclidean similarity between two different environments.

\subsection{Free Energy Construction and Clustering}

The final stage of the unsupervised learning pipeline, involves examining the free energy landscape(FES) of the sampled conformations. Specifically, in this work, we seek to construct the high-dimensional thermodynamic landscape that connects the water environments in bulk water to those around the ions. To achieve this, we first apply the Point Adaptive K-nearest neighbors(PAk) estimator\cite{PaK} to compute the free energy of each point in the dataset. The PAk estimator uses statistical approaches\cite{Knn_1979} for estimating the local density($\rho$) of points in high-dimensional space to the computation of free energy. The point-dependent free energy is derived from the local density using the equation:

\begin{equation}
    F_i=-k_{B}log(\rho_i)
\end{equation}

In the PAk framework, the free energy estimation is based on determining the volume of the hypersphere around each point within which the density can be considered constant (around a given statistical confidence interval). Importantly, these volumes are computed on the manifold where the data reside, rather than in the feature space, and therefore circumvents the need to construct collective variables defining this manifold. However, the point-free energies need as input, the ID of the manifold which can be calculated for example, by the two-NN estimator described before. The PAk estimator offers two critical advantages: it provides an estimate of the statistical uncertainty associated with the free energy, and it allows for constructing free energies in their intrinsic high-dimensional space, thereby mitigating the risk of information loss. 

Once the point-dependent free energies are determined, we apply the Advanced Density Peak clustering algorithm (ADP)\cite{automatic_topography}, which is a refined version of the original Density Peak clustering algorithm\cite{DP}, to group similar
environments and identify relationships between the resulting clusters. 
In this procedure, each cluster is centered on a specific point, corresponding to a free energy minimum and the level of coarse-graining is determined by a single parameter $Z$. Indeed, the merging of two putative clusters $a$ and $b$, occurs if the following condition is satisfied:

\begin{equation}
F_{ab} - F_{a} < Z (\epsilon_{F_a}+ \epsilon_{F_{ab}} )
\label{Z_equation}
\end{equation}

Here, $F_{a}$ is the free energy at the center of cluster $a$,  with $\epsilon_{F_a}$ denoting its associated uncertainty. Similarly, $F_{ab}$ is the free energy of the saddle point connecting cluster $a$ and $b$ and $\epsilon_{F_{ab}}$ is its uncertainty. By lowering $Z$, it is possible to obtain a more detailed representation of the free energy landscape as it resolves more of its corrugations. However, care must be taken to balance the level of detail with the statistical significance of the identified clusters. On the other hand, increasing the $Z$ too much leads to a more coarse-grained view of the FES and can subsequently wash out important details. Another important aspect of ADP is that it also provides a hierarchical view of the clusters (referred to as the Dendrogram representation) that, as we will later see, can immediately be interpreted as the topography of the free energy landscape.

\section{Results}
\label{sec:results}

Our approach in this work, is to give a deeper dive into the challenges and opportunities in using unsupervised learning approaches to understand physical phenomena. Here, we specifically focus on the water structure of electrolyte solutions generated from molecular dynamics simulations. The following section will explore three main sub-themes: 1) the selection of hyperparameters needed to build SOAP descriptors, 2) the extraction and interpretation of the intrinsic dimension (ID) of the high-dimensional SOAP features, and 3) the construction and interpretation of high-dimensional free energy landscapes. In all steps, we will highlight the importance of physical and chemical knowledge in guiding the interpretability which is often a challenge, as well as in guiding the selection of hyperparameters.

\subsection{Construction of the SOAP Features}

Our overarching goal is to be able to characterize the water environments around ions in an agnostic way using data harvested from classical molecular dynamics (MD) simulations of several divalent and monovalent cations (calcium, magnesium, lithium, potassium, rubidium, cesium) and anions (chloride, fluoride, bromide, and iodide) in liquid water.  We select SOAP coordinates\cite{SOAP_2013} as features to describe the water environments around the ions. This choice reflects our focus on the structural and thermodynamical properties of solvation: SOAP provides a robust, high-resolution, and symmetry-invariant representation of local solvation structures, enabling  an unsupervised analyses of water environments around different ions. As seen earlier, the SOAP descriptor around a specific chemical species involves expanding the atomic density in terms of radial and spherical harmonic basis functions. Defining the relevant environment surrounding each ion involves several key physical considerations (earlier described), that translate into a set of specific hyperparameters that need to be selected.

Firstly, the SOAP features are centered on the ion of interest with Gaussians placed on all the oxygen and hydrogen atoms belonging to water molecules located within a cutoff radius, $R_{cut}$, contributing to the SOAP density. 
For water, used as a reference system, the coordinates are calculated by randomly selecting an oxygen atom at each simulation frame. Gaussians are then placed on all nearby oxygen and hydrogen atoms, except for the two hydrogens that are covalently bonded to the selected oxygen atom. By constructing the SOAP environments in this manner, water environments in neat water can be directly compared to those around an ion.

One obvious challenge in constructing the SOAP descriptor is what criterion to use for the radial cutoff of the environment around a water molecule or an ion ($R_{cut}$). Our focus is specifically on how ions perturb the local structure of water in their immediate vicinity (the first solvation shell). The typical approach to determine the $R_{cut}$ parameter is examining the pair-correlation function between a pair of particles in this case for example, the ion and surrounding water. Since we were interested in comparing water environments from different ions, we initially attempted to use the same cutoff for all ions. As we will see later, this choice did not adequately allow us to resolve important physical differences between the water network around the different ions and we therefore constructed cutoffs that were specifically designed for each ion separately. 

\begin{center}
    \begin{figure}[H]
       \includegraphics[width=\columnwidth]{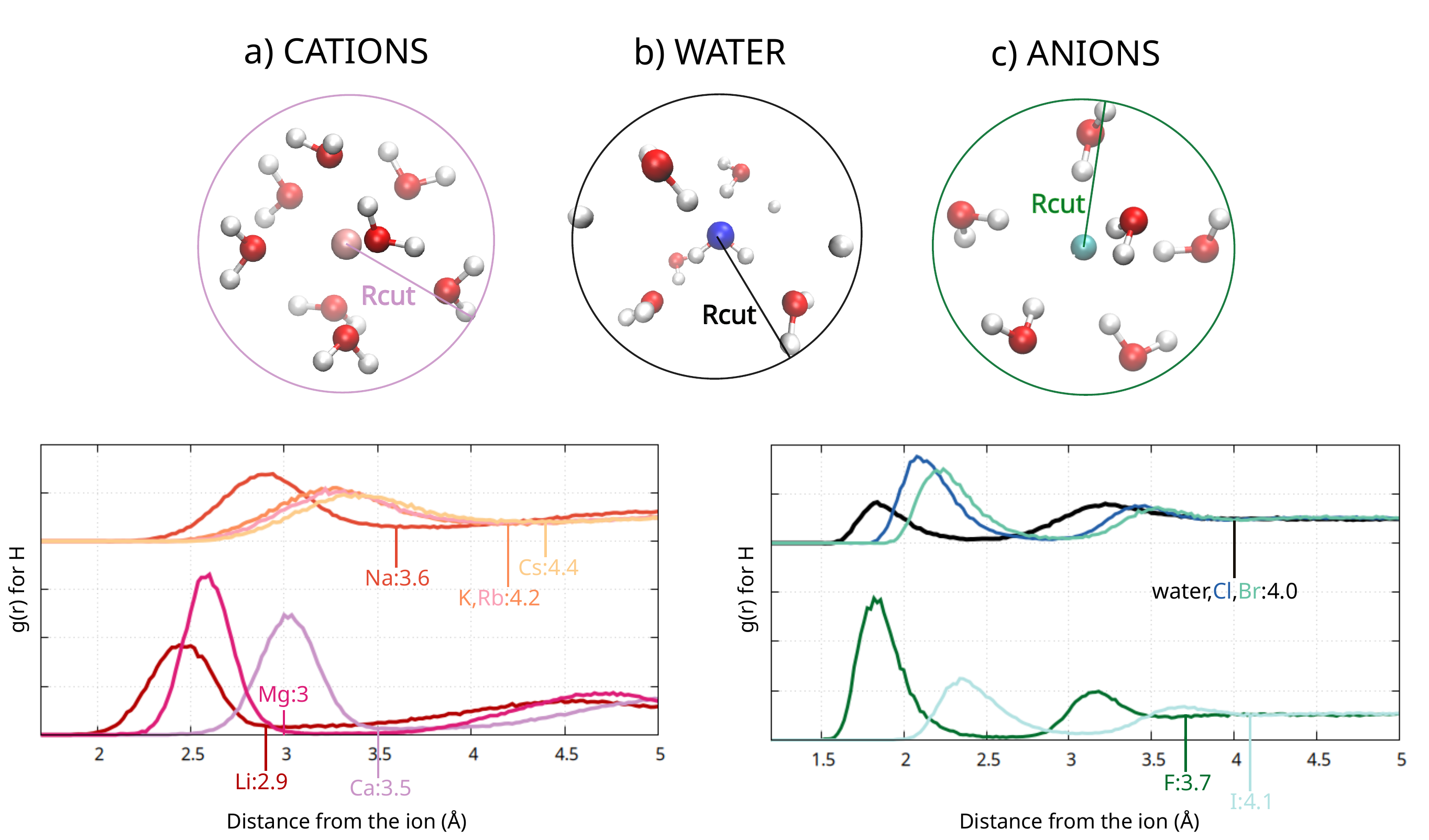}
	       	\caption{Upper panels: representative snapshots show how the first solvation shells are built for cations(panel a), bulk water(panel b) and anions(panel c). Lower panels: radial distribution functions of hydrogen atoms of water molecule around cations(left panel) and both around anions and around a central oxygen atom of bulk water(right panel).  }
		      \label{first_shells}
    \end{figure}
\end{center}

The representative snapshots in the upper panels of Figure \ref{first_shells} illustrate how the first shell is constructed for cation, bulk water and anion respectively. In the case of a cation, water molecules tend to be less oriented as the two water protons are placed further way and on average tend to be at similar distances from the ion (Figure \ref{first_shells}a). On the other hand in an anion, water molecules are more orientationally polarized due to the fact that one proton tends to be closer to the negative charge of the ion compared to the other (Figure \ref{first_shells}c). A water molecule can both accept and donate H-bonds (Figure \ref{first_shells}b) and therefore exhibits characteristics of both the cationic and anionic environments. For the cations, we thus used the position of the first minimum associated with the ion-proton radial distribution function (g(r)) while for water and anions we used the second minimum. Figure \ref{first_shells} illustrated the g(r) for water, cation and anion along with the values of the ion-specific cutoffs that are employed.

Beyond the radial cutoff, an additional critical SOAP parameter is the width ($\sigma$) of the Gaussian functions, which defines the chemical resolution at the relevant length scales in the system. Some of the earlier studies using SOAP to characterize liquid water\cite{Adu_liquid_water2022,Bingqing_ice_phases_2020,capelli2022_ice_environments}, adopted a $\sigma$ of 1$\r{A}$, motivated by the observation that the length of a hydrogen bond is roughly three times larger. 
However, as shown by Pozdnyakov and co-workers\cite{Gabor_sigma_vs_cardinalita_2021}, there exists an interplay between the size of the basis set and the chosen $\sigma$, which can influence the description of atomic environments. In more recent work, we have demonstrated, through information theoretical techniques, that in order to capture important chemical details of the hydrogen-bonding network, a $\sigma$ smaller than 1$\r{A}$ needs to be employed albeit the challenge of dealing with SOAP features with larger cardinality\cite{eddy_samestory}. For this work, we settled on a $\sigma$ of 0.4\r{A} which, as we will see later, is small enough to capture trends that we are after.

While determining the $R_{cut}$ and $\sigma$ can be relatively easily justified from physical and chemical intuition, the choice of the number of radial basis functions ($N_{max}$) and the number of spherical harmonics ($L_{max}$) is much more challenging. This in part originates from the fact that a priori it is not known what shape and orientation of the relevant orbitals of the spherical harmonic expansion are needed to correctly describe the water environment of an ion. To address this challenge, we next turn to the ID as a diagnostic for selecting $N_{max}$ and $L_{max}$.

\subsection{$N_{max}$ and $L_{max}$ Selection from Intrinsic Dimension}   
\label{Nmax_Lmax_choice}

The ID of a dataset coming from an atomistic simulation quantifies the number of independent degrees of freedom needed to describe the fluctuations in the system. The ID has been shown to provide a quantitative manner of probing changes in a wide class of different systems ranging from spin-models\cite{alex_ID_ising, PRXQuantumID} to proteins\cite{allegra_michele_IDvariabile} or even Neural Networks representations\cite{ansuini2019intrinsic,valeriani2024geometry}. In our case, we use the ID as a proxy of the extent to which an ion induces order or disorder relative to the case of water environments around a neutral water molecule. Broadly speaking, larger IDs imply less correlations between the different degrees of freedom involved while smaller IDs would suggest higher correlations and subsequently more order\cite{basile2024intrinsic}. The physical intuition would be that the electric field from the charge (cation/anion) is larger than that from a neutral water (dipole) and therefore the presence of the ions would increase the correlations within the hydrogen bond network. 

\begin{center}
    \begin{figure}[H]
       \includegraphics[width=\columnwidth]{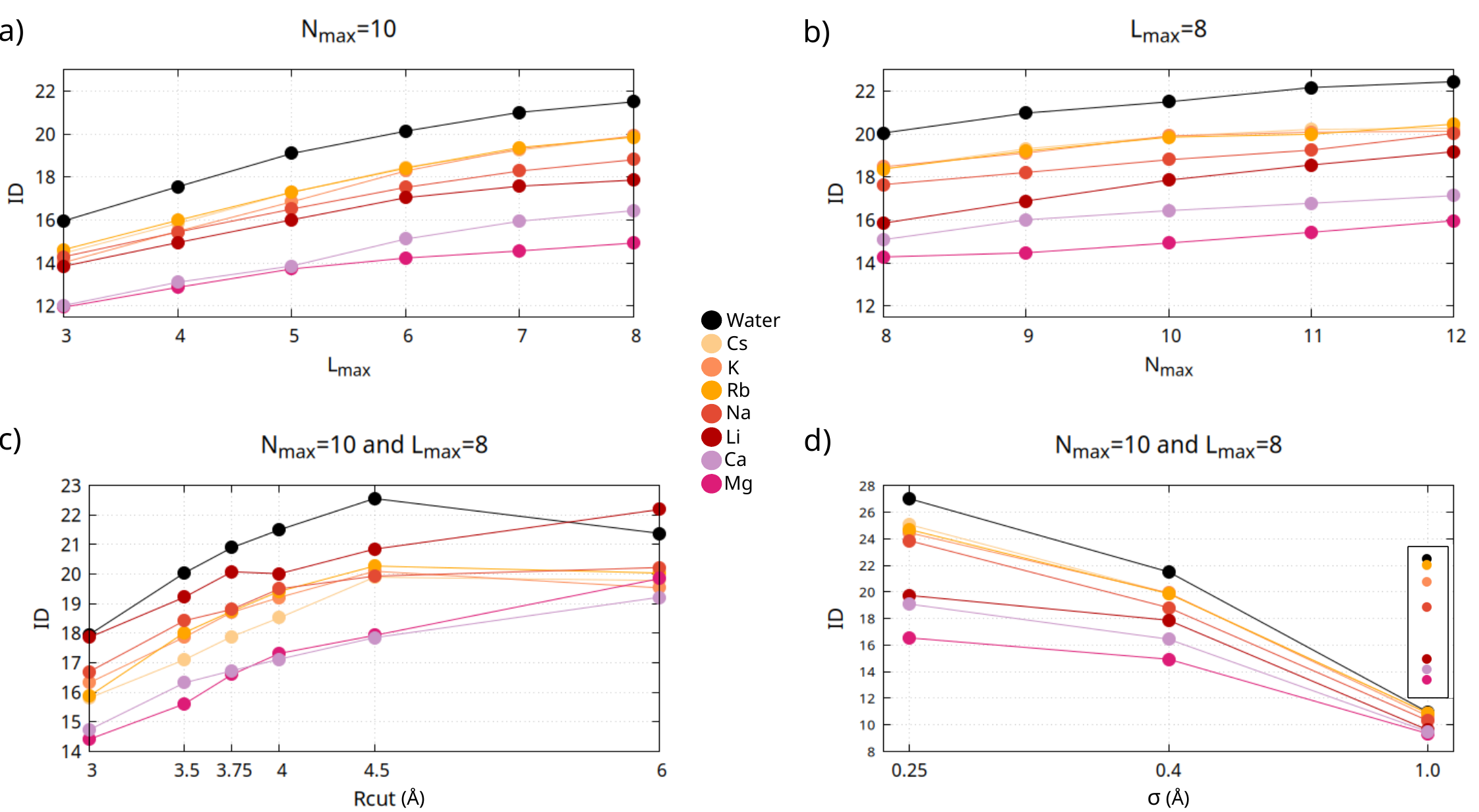}
	       	\caption{IDs of cations and of bulk water as a function of the SOAP parameters. In panels a,b,d the $R_{cut}$ defining the first solvation shells are defined based on the ion-proton g(r). Panel a: IDs as a function of $L_{max}$, fixing $N_{max}=10$. Panel b: IDs as a function of $N_{max}$, fixing $L_{max}=8$. Panel c: IDs as a function of a single cut-off ($R_{cut}$) defining the first solvation shells, fixed to be the same for all ions. Panel d: IDs a function of sigma parameter, fixing $N_{max}=10$ and $L_{max}=8$. }
		      \label{scanning3}
    \end{figure}
\end{center}

Initial analyses indicated minimal variations in the IDs of the anions; we thus focused on examining the sensitivity of the ID of water surrounding the cations, to the choice of the $N_{max}$ and $L_{max}$. 
Notice that, in all panels of Figure \ref{scanning3}, error bars (see the Methods section for details on how they are calculated) are omitted as they fall within the radius of the data points. Figure \ref{scanning3}a illustrates how the ID evolves as a function of $L_{max}$ at a fixed $N_{max}=10$, while Figure \ref{scanning3}b shows the ID dependence on $N_{max}$ at a fixed $L_{max}=8$. In both cases, the ID grows as a function of $L_{max}$ and $N_{max}$, reflecting a larger number of degrees of freedom needed to describe the SOAP space as the basis-set dimension expands. In addition, there appears to be a ranking of the IDs with water being the highest followed by monovalent cations and then finally the di-valent cations being the lowest. This hierarchy is captured across all parameter sets tested and reflects the enhanced correlations present between the water molecules as the ionic charge increases. 

Looking more carefully at Figure \ref{scanning3}a, we observe that for $L_{max}\ge6$, an additional sub-ranking emerges among ions of the same charge. Indeed among the divalent group, magnesium exhibits a lower ID than calcium while among the monovalent one lithium has the lowest ID followed by sodium and finally, potassium-rubidium-cesium whose IDs, are indistinguishable from each other. Figure \ref{scanning3}b confirms that this sub-ranking persists at sufficiently large values of $L_{max}$ (i.e. $L_{max}=8$) across the entire range of $N_{max}$. Based on these observations, we selected $N_{max}=10$ and $L_{max}=8$ for the subsequent analyses: these parameters values capture the refined IDs hierarchy. In the following section, we provide the physical and chemical rationale underlying these trends.

In the results presented in the first two panels of Figure \ref{scanning3}, ion-specific cutoffs were employed. One may argue that the ion specific choice of the radial cutoffs could introduce inconsistencies in the description of the environments, for example when comparing the IDs. To assess the impact of using different cutoffs, we investigated how the previously observed ID hierarchy changes when using a single radial cutoff for all ions, fixing $N_{max}=10$ and $L_{max}=8$. Specifically, Figure~\ref{scanning3}c shows the IDs of all cations as a function of the radial cutoff, which is set to the same value for each system.
Within the 3.5-4.5\r{A} range, the coarse-grained ranking of IDs is preserved, with water displaying a higher ID than the monovalent cations, which in turn exhibit higher IDs than the divalent cations. However, the finer sub-ranking among ions of the same charge is no longer reproduced. For example, lithium appears to have a higher ID than all other monovalent cations. 
This discrepancy can be attributed to lithium’s smaller ionic radius and higher charge density, resulting in a more compact first solvation shell compared to other monovalent cations (see the g(r) of Figure \ref{first_shells}). Consequently, under a fixed radial cutoff, let' s say $R_{cut}=3.75$\r{A}, the volume probed for the other monovalent cations spans primarily the first solvation shell, whereas for lithium, atoms from the second solvation shell are already included, thereby increasing the complexity of its local environment and raising its ID. These observations demonstrate that using an ion-specific radial cutoff is necessary to ensure a fair comparison of the atomic environments across different ions.

Finally, the ordering of the IDs shows relatively little dependence on the choice of Gaussian width. Figure \ref{scanning3}d illustrates the IDs of all the systems, fixing $N_{max}=10$ and $L_{max}=8$, at three Gaussian widths: $\sigma=0.25, 0.4, 1.0$\r{A}. In all cases, the overall ID ranking is preserved; however, at $\sigma=1$, all the IDs collapse into a narrow range. This collapse arises because a larger $\sigma$ increases the coarseness of the space description, thus masking finer structural differences among the different ionic environments.

To conclude this part of our analysis, we have shown that the ID provides a single number that quantifies the correlations within the water network surrounding each ion. Although the precise value of the ID depends on the specific construction of the SOAP descriptors, a robust hierarchical ordering of the IDs emerges across a broad range of SOAP parameters, provided that the radial cutoff is adapted to encompass each ion’s first solvation shell. The following section will delve into the physical interpretation of this observed ID hierarchy.

\subsection{Physical and Chemical Interpretability of ID}

\begin{figure}[H]
        \includegraphics[width=\columnwidth]{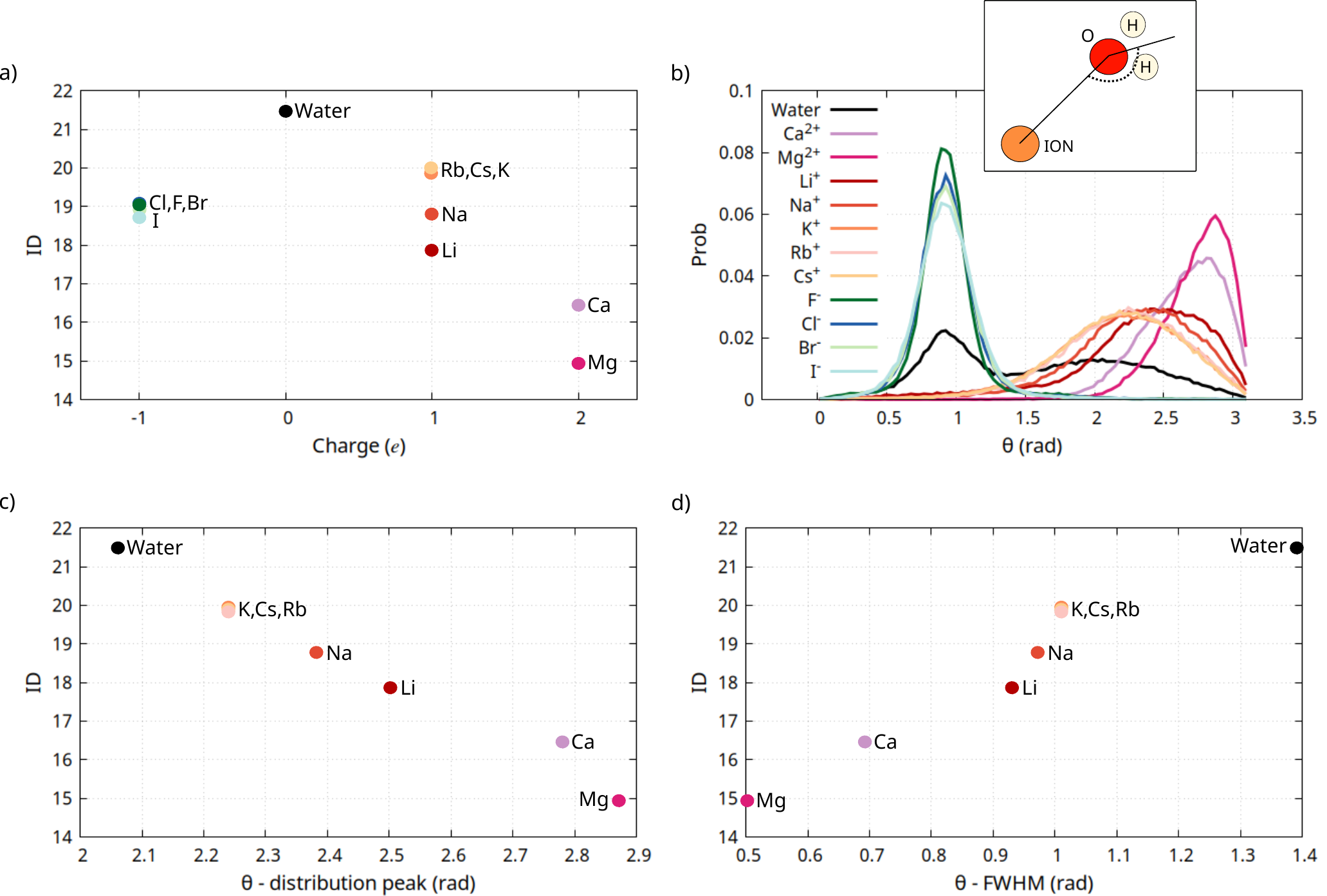}
		\caption{Panel a: ID of all studied systems(anions, cations and bulk water) as a function of the ionic charge(measured in $e$). Panel b: normalized distributions, for each ion, of the dipoles orientations of water molecules belonging to the ionic first solvation shell. The dipole orientation is defined by the $\theta$ angle, represented in the insert. Panel c: ID versus the peaks positions of angular distributions of panel b. Panel d:ID versus the peaks FWHM of angular distributions of panel b.}
		\label{IDs_dipole_orientations}
\end{figure}

One of the open and important challenges in the use of ML based approaches is that of interpretability. In the current context, one of the outcomes of the ID analysis is the presence of a ranking which is consistent with our intuition that the presence of a charge reduces the number of independent degrees of freedom of the water hydrogen-bond network. These findings are summarized in Figure \ref{IDs_dipole_orientations}a - the IDs of all ions are lower than the one of bulk water, with anions exhibiting very similar values. In contrast, the cations display a clear separation between monovalent and divalent species, and, as noted in the previous section, there is a further subranking among cations of the same charge. In what follows, we focus on providing a physical interpretation for these ID ranks.

Cations and anions induce very different orientational effects on the surrounding water molecules. Specifically, negative charges tend to orient the dipoles stronger because they can attract the hydrogen atoms of the water molecules. To quantify this effect and to examine if this is reflected in the ID, we examined the orientational distribution of dipoles of water molecules belonging to the first solvation shell of each ion. The dipole orientation is determined by the angle formed by the radial vector connecting the central ion to the oxygen atom of the water and by the dipole of the water molecule (we denote this angle $\theta$ - see the insert of Figure \ref{IDs_dipole_orientations}b). In Figure \ref{IDs_dipole_orientations}b we illustrate these orientational distributions for neutral water and all the ions.  For the case of neutral water, the distribution has two peaks which essentially correspond to water molecules on the accepting (larger value of the angle) and donating side (lower value of the angle). Water dipole orientations around the anions or cations are exclusively characterized by a single peak. Starting with the anions, we observe that all of them maintain the same peak position at the same value as that associated with water molecules donating hydrogen bonds to the negative charge on the oxygen atom and display only a very subtle change in the width of the distribution. On the other hand for the cations, we observe more significant changes in both the peak positions as well as the widths.

Figure \ref{IDs_dipole_orientations}c and \ref{IDs_dipole_orientations}d illustrate the correlations that emerge between the magnitude of the ID and the peak position/width of the angular distributions. Panel c shows a striking anti-correlation between the peak position of cations (and of the cationic peak of water) and the ID of the corresponding environment: the more the water dipoles are aligned to the ion's electric field, the lower is the ID of the corresponding ionic environment. On the other hand, the width of the distributions as determined from the full-width at half maximum (FWHM) is positively correlated with the ID which originates from the fact that the larger fluctuations of the water dipole imply a more disordered hydrogen-bonded network. These results suggest that the number of degrees of freedom of cationic environments is basically dictated by the capability of the ion to orient and constrain water dipoles. In the case of the anions, the indistinguishability among the IDs of different ions is reflected in very similar peak positions of the $\theta$ distributions. Moreover, it appears that the variations in the distributions widths are also rather too small to be manifested in the IDs.

\begin{figure}[H]
\center
        \includegraphics[width=\columnwidth]{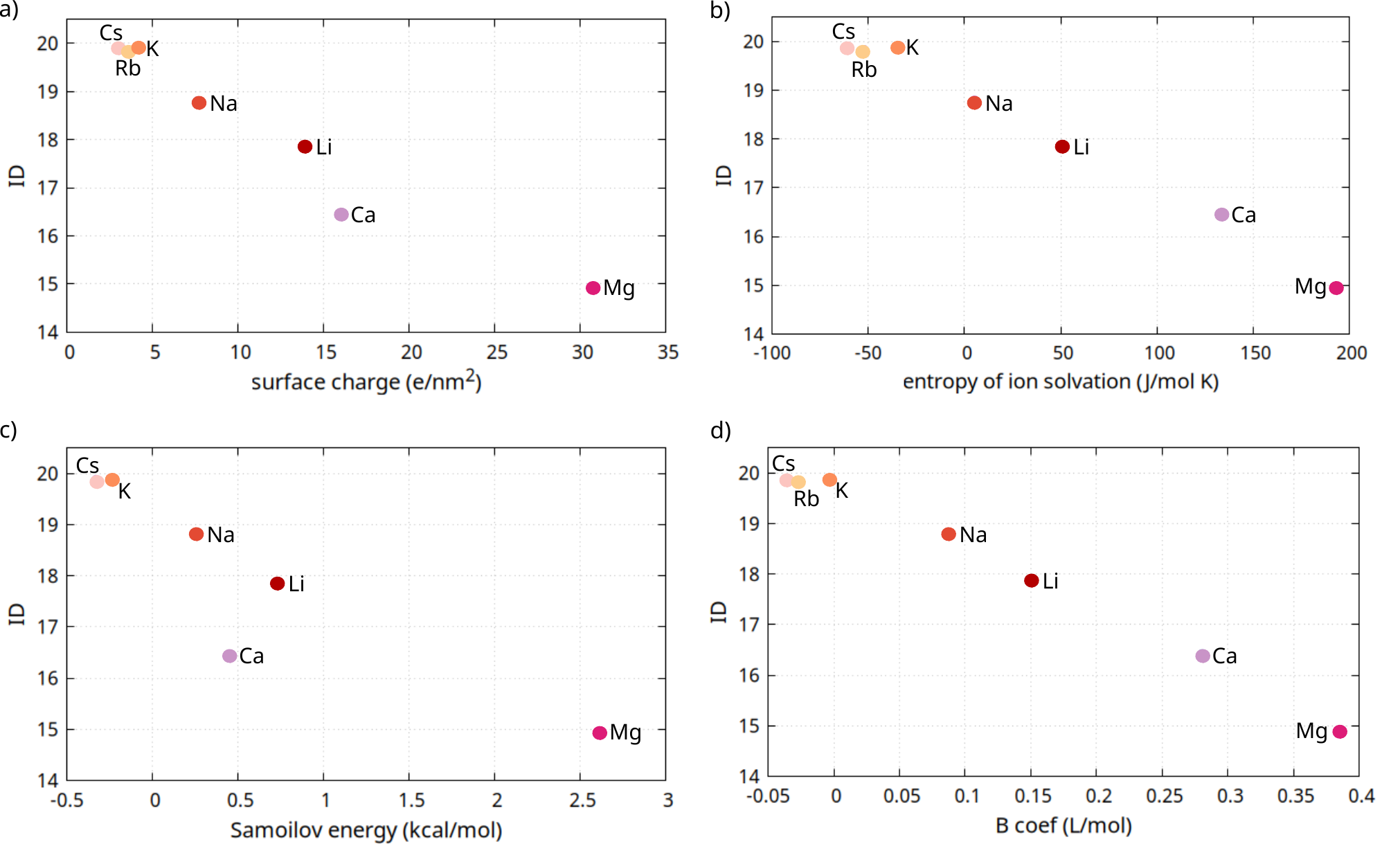}
		\caption{Correlations between cationic IDs and experimental properties of the corresponding ions. Panel a: ID versus the ionic surface charge, measured as charge/$4\pi r^2$, where the charge is measured in $e$ units and $r$ is the ionic radius from Ref. \citenum{shannon1976_ionic_radii}. Panel b: ID versus entropy of ions solvation (from Ref.\citenum{krestov_thermodynamics_1990}). Panel c: ID versus Samoilov activation energy (from Ref. \citenum{samoilov_1957}). Panel d: ID versus Jones-Dole B coefficient (from Ref. \citenum{Bcoef_1957}).    }
		\label{exp_correlations}
\end{figure}

As eluded to earlier, the response of water to the presence of ions has been long-discussed in the context of the Hofmeister series which reflect a ranking of ions with different tendencies to either salt-in or out proteins due to their apparent structure making or breaking character. Over the last decades, numerous empirical parameters have been derived or inferred from experimental measurements. In search for some more physical origins of the ID, in Figure \ref{exp_correlations} we summarize the correlations that are observed between the ID and parameters such as the surface-charge of the ion, the entropy of ion solvation\cite{krestov_thermodynamics_1990}, the Jones-Dole coefficient\cite{Bcoef_1957} and finally the Samoilov energy\cite{samoilov_1957}. Curiously, the general trend observed is that the ID is anticorrelated with all four parameters. 

Figure \ref{exp_correlations}a shows the trends that are observed between IDs of cations and their surface charge, where the latter is defined as the ionic charge (in electron units) divided by the surface area of a sphere (using the ionic radii selected from Ref. \citenum{shannon1976_ionic_radii}). The anticorrelation observed between these two variables along with what we observed in Figure \ref{IDs_dipole_orientations}, provides a consistent physical basis of the ID ranking of all cations: a larger surface charge density allows the ion to fix the dipoles of surrounding water molecules which in turn results in a more correlated hydrogen bond network and therefore a smaller ID.
Panels b and c show that the degree of order of the water environment has strong effects on the physical behavior of the corresponding system and that this is reflected in the magnitude of the ID. In detail, panel b shows an anti-correlation between the IDs and the entropy of ion solvation (data from Ref. \citenum{krestov_thermodynamics_1990}): the lower is the ID, the larger is the entropy of solvation. 
Interestingly, the three larger ions, namely cesium, rubidium, and potassium, which have a negative entropy of solvation, consistently exhibit larger IDs. An analogous behavior is also observed in the Samoilov activation energy (Figure \ref{exp_correlations}c - data from Ref. \citenum{samoilov_1957}), defined as the energy required to remove a water molecule from the ion’s first solvation shell relative to removing it from that of another water molecule. All the ions, except for calcium, show similar trends in ID to those observed in the entropy of solvation. It should be stressed that this quantity can only be inferred by combining experimental data with theoretical models.

Finally, Figure \ref{exp_correlations}d illustrates a clear anti-correlation between the ID and the Jones-Dole B coefficient (data from Ref. \citenum{Bcoef_1957}). This coefficient is a parameter in the empirical Jones-Dole equation\cite{jones_dole_1929} describing the relative viscosity of electrolyte solutions as a function of solute concentration and has been used extensively in the literature to infer the structure making/breaking character of ions (see Discussion and Conclusions section later). Taken together, these correlations confirm our intuition on how the ID of water environments, built on SOAP features, captures important information on the correlations within the hydrogen-bond network which is then mirrored in different thermodynamic properties.

\subsection{Constructing the Free Energy Landscape}

Following the ID analysis of the environments surrounding individual ions, we are now in a position to construct the high-dimensional free energy landscape in SOAP space. 
To achieve this, we first applied the PaK estimator\cite{PaK} to compute the free energy at each point of the combined dataset of SOAP features from bulk water, cations and anions (this data set now consists of 120000 SOAP features arrays).
As illustrated earlier, the different ionic environments shown in the previous section, are characterized by different IDs; however, to allow a consistent comparison of free energy values across all environments using PaK, a single ID representative of the entire dataset must be selected (as detailed in the Methods section). The Two-NN estimator determined an ID of $\sim$19 for the dataset, which lies within the range of IDs observed for the individual ionic environments. Subsequently, we present the results obtained using an input ID of 19 for PaK, while showing in Figure S1 in the Supplementary Information (SI) that the results remain essentially unchanged within a broad ID range ($16\leq$ ID $\leq 22$).

After having obtained the free energy value at each point, we utilized the Advanced Density Peak clustering algorithm(ADP)\cite{automatic_topography} to group similar environments and identify relationships between the resulting clusters. Each cluster is centered around a specific point representing the free energy minimum for all elements within the cluster and the saddle points among different minima are then used to determine the free energy barriers. Note that the term ‘Free energy Landscape’ is here used in a broad sense: points in the dataset are not temporally connected and there are no transitions between the identified clusters in the simulations since they come from independent trajectories. However, our SOAP descriptors are built looking only at the water environments around a water molecule or ion (excluding the central species) – it is thus possible to compare the free energy landscape in SOAP-space of these environments, identifying their overlapping-similarity.
The clustering process depends on a single parameter ($Z$), which determines the level of coarse-graining in defining the clusters: the larger the $Z$ value, the higher the free energy barrier between two different clusters that will subsequently not be merged (see Equation \ref{Z_equation}). 

\begin{figure}[H]
\onecolumn
        \includegraphics[width=\columnwidth]{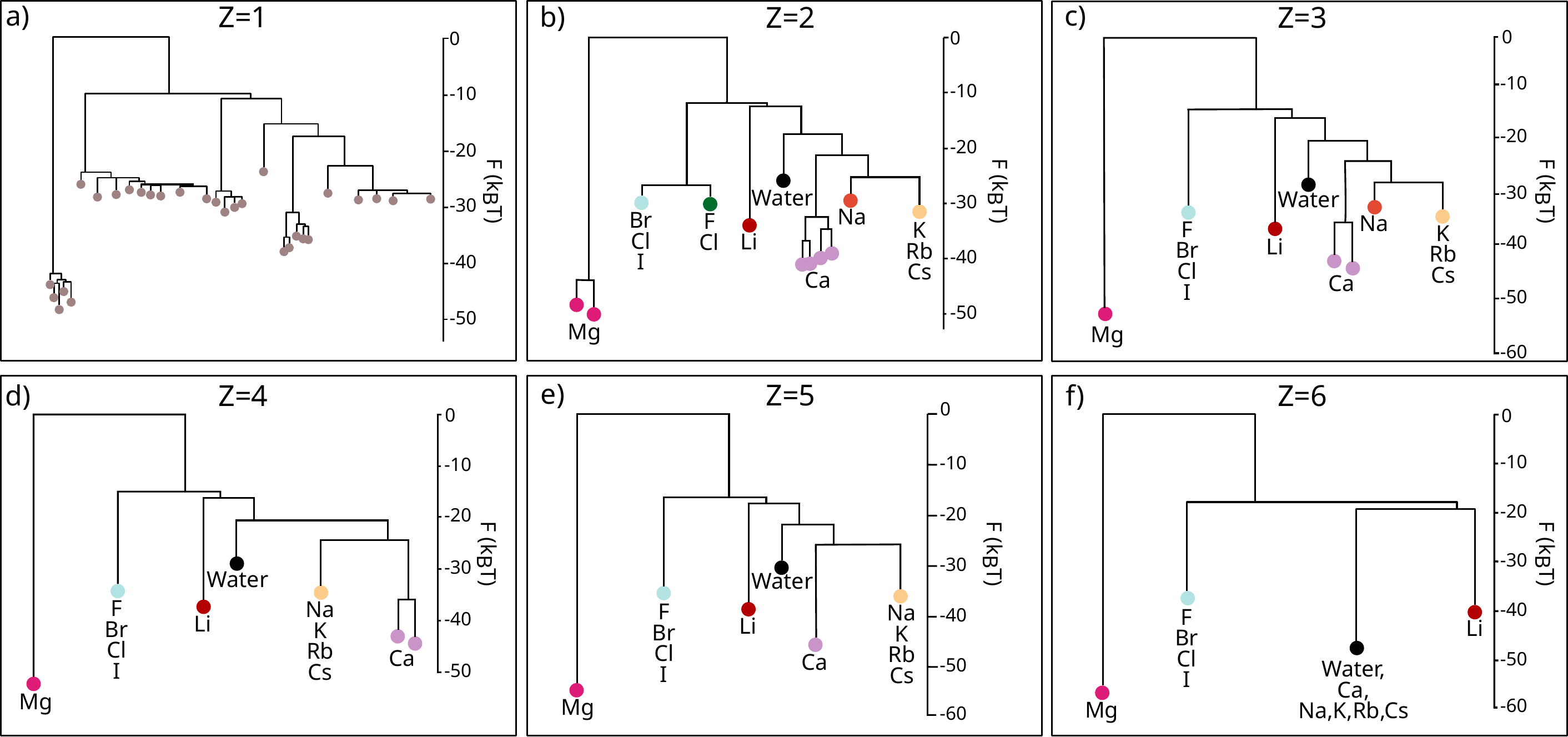}
		\caption{Free energy landscapes of the dataset comprising all ionic water environments and bulk water environments, for six values of the merging parameter of the ADP clustering algorithm ($Z$ equal to 1,2,3,4,5 and 6). Landscapes are presented using the dendrogram representation.}
		\label{comparing_Z}
\end{figure}

Figure \ref{comparing_Z} illustrates the landscapes obtained for six different $Z$ values (1,2,3,4,5 and 6) using a dendrogram representation.
To minimize noise, clusters whose population is smaller than 0.01\% of the total dataset are excluded from the analysis.
In the case of $Z=1$, no labels are assigned to the clusters, as a total of 29 clusters are identified, making a detailed examination of each individual cluster impractical.
The figure demonstrates how reducing the coarse-graining level ($Z$) leads to an increased number of clusters, offering a more detailed description of the landscape. 
Lithium and magnesium clusters are the first to emerge as distinct single-ion clusters from the multi-ionic ones at $Z = 6$, separated by high free energy barriers($\geq 20k_{BT}$). As we will discuss later, this is a consequence of a well defined geometrical arrangement of water molecules around those specific ions. 
Reaching $Z = 2$, single ionic clusters, such as the calcium and magnesium ones, start to get separated. The barriers are however rather small and cannot be physically rationalized. We thus focus our analysis on the landscape obtained at $Z = 3$, which offers an good balance of resolving the different water environments around the ions but at the same time avoiding unphysical partitioning of the free energy landscape.

\begin{figure}[H]
\onecolumn
        \includegraphics[width=\columnwidth]{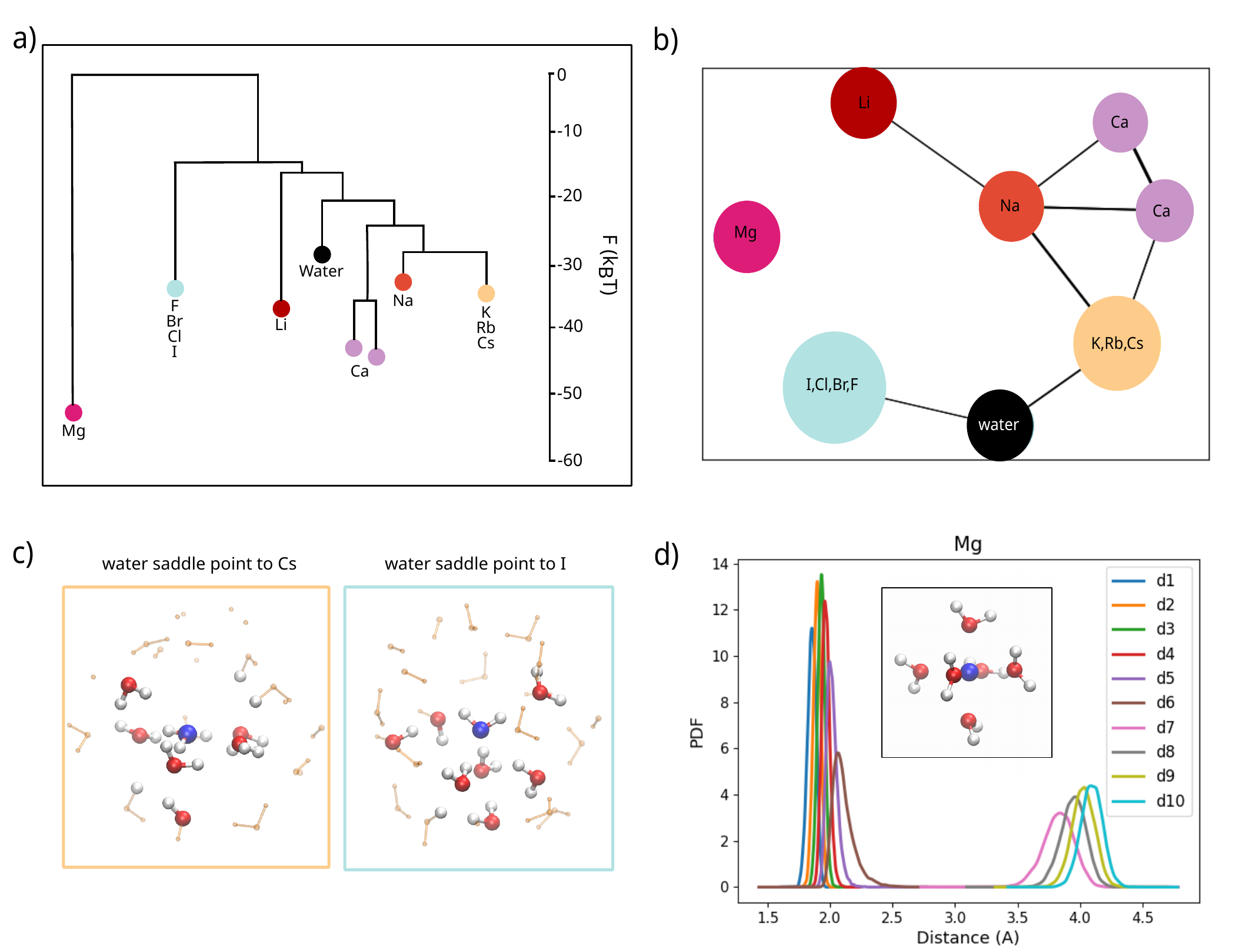}
		\caption{Free energy landscape obtained at $Z=3$: physical insights. Panel a: free energy landscape representation through a dendrogram. Panel b: free energy landscape representation through a 2D projection. Panel c: Left - saddle point of the water cluster, connecting it to the cesium cluster. Right - saddle point of the water cluster, connecting it to the iodine cluster. Panel d: Normalized distributions of the first ten magnesium-oxygen distances, evaluated at each snapshot of the $1\mu s$ simulation. In the insert a representative snapshot illustrates the octahedral geometry of water molecules(of the first solvation shell) around magnesium ion. }
		\label{FES_Z3}
\end{figure}

Figure \ref{FES_Z3}a shows the dendrogram corresponding to the free energy landscape, fixing $Z = 3$. The first striking observation is that the unsupervised approach operating in the space of SOAP coordinates effectively partitions water ionic environments with many clusters corresponding to specific ionic species. Many of the ions thus distinctly modify the water environments creating thermodynamically stable states that are however, connected to other ionic species through fluctuations of the water hydrogen-bond network. There are however, two notable exceptions: the cluster containing all anions and the cluster comprising potassium, rubidium, and cesium environments. This feature is consistent with previous observations on the similarity of the IDs as well as in the dipole orientational distributions (see Figure \ref{IDs_dipole_orientations}a and \ref{IDs_dipole_orientations}b). In essence, the fluctuations of the relevant degrees of freedom involving the solvent around the ions create similar environments. This type of variability is reflected also in Figure S2 of SI which shows the distributions of the distances of the nearest ten water molecules to each ion. A huge overlap is indeed observed in all the distances distributions of anions, potassium, cesium and rubidium.


To better understand the hierarchical structure of the dendogram and the connections between the different basins we provide in Figure \ref{FES_Z3}b a complementary representation. Here specifically, the circle sizes indicate cluster populations, and connections between clusters are drawn only when a saddle point exists between them. The distances between clusters are optimized to be, as closely as possible, proportional to the free energy barriers. From this figure, it emerges that, although the dendrogram visually places the bulk water cluster among cationic clusters, it is directly connected only to two multi-ionic clusters: the one containing anions and the one containing potassium, rubidium and cesium. The physical implications of this are that fluctuations in bulk water without any explicit ions tend to create through the formation of defects, environments that resemble those around cations and anions. This is best seen through the snapshots illustrated in Figure \ref{FES_Z3}c for the water saddle point close to cesium and iodide(left and right panels) where the environment around the central water molecule is shown. Atoms in bold are within the first solvation shell. In both cases, we observe the creation of defective water environments involving large asymmetries in the accepting/donating nature of the central water molecule and the creation of cavities. These transient local fluctuations in bulk water can thus mimic either a cationic- or anionic- like solvation environment.

The divalent cations (magnesium and calcium) appear to display some interesting features from our free energy analysis. Starting with magnesium, we observe from Figure \ref{FES_Z3}a and \ref{FES_Z3}b, that it is completely separated from all the other clusters (which appears to be robust for all $Z$ values) - in other words, there are no saddle points between it and any other cluster. Magnesium, as the cation with the highest charge density, has a first solvation shell comprising six water molecules arranged in an octahedral geometry (as depicted in the insert of Figure \ref{FES_Z3}d). The solvation environment around magnesium is a rigid one without deviations from the ideal octahedral geometry: there are no feasible pathways connecting magnesium’s local environment with those of other ions, resulting in significant barriers. This rigidity is demonstrated both by the low fluctuations of the magnesium-oxygen atoms distances (Figure S2 in the SI) and of the water dipoles (Figure \ref{IDs_dipole_orientations}b).  In the case of calcium, we consistently observe that it partitions into two subclusters, separated by a significant barrier ($\geq 5k_{B}T$). The reason for this subdivision is the presence of two well distinct possible geometries of water environments around calcium: in one case, the ion is surrounded by seven water molecules and in the other by eight, as shown from the distributions  of Figure S3 in the SI. 


\section{Discussion and Conclusions}

In this work, we extended the application of an unsupervised protocol recently developed in our group for the study of different aqueous systems to investigating the behavior of water around ions. One of the key objectives here was to investigate thoroughly the technical aspects inherent to each step of the procedure, highlighting the challenges for new-comers to the field, with an important emphasis on the importance of selection of hyperparameters when performing unsupervised learning. 

To this end, we employed SOAP descriptors given their suitability in capturing with high-resolution, structural and thermodynamic features of liquids\cite{solana_zundeig,Adu_liquid_water2022,eddy_supercooled_2024}.The development and application of local atomic descriptors like SOAP to molecular systems is currently a very active area of research. Recently, Pavan and co-workers have proposed alternative SOAP-based  descriptors combining structural and temporal information\cite{timesoap_2023_Pavan,lens_2023_Pavan}, which have proven effective in detecting sparse but dominant fluctuations in liquids structure and in comparing the physical information at different length-scales.

We have shown that the construction of ML descriptors involves a step requiring careful evaluation. A key challenge in employing ML descriptors for structural analysis lies in the choice of the hyperparameters and subsequently in the complexity of physically interpreting the feature vector. Physical intuition and perhaps heuristics can be crafted to perform this step however, this may not always be possible. We show that the intrisic dimension (ID) provides one manner in which to probe the complexity of the network and to guide the choice of hyperparameters associated with the SOAP descriptor. We find that radial cutoffs that are ion-specific need to be selected - the implications of this in the design of neural-network potentials for electrolyte solutions would be an interesting avenue to explore in the future. 

Investigating the ID of the water around the ions, we find that they all tend to lower this value compared to the neutral bulk case. This indicates that the presence of an ion always enhances the correlations within the water hydrogen bond network. In this sense, one may argue that all single ions are structure makers. This is further reinforced by the trends we observe between the IDs and thermodynamic observables such as the entropies of solvation and the Jones-Dole B coefficient. The latter is a parameter in the empirical Jones-Dole equation describing the relative viscosity of electrolyte solutions as a function of solute concentration: ions with positive B coefficients correspond to \emph{structure makers} in the Hofmeister series, whereas ions with negative B coefficients are the \emph{structure breakers}\cite{Bcoef_1957,Bcoef_1959}.  The connection between the ID and the B coefficient is at least suggestive that despite being empirically derived, likely encodes important microscopic information on the water hydrogen-bond network.

Finally, the free energy analysis in high dimensions provides a nuanced manner of comparing the solvation environments coming from different ions. The water environments in terms of their orientational structure around all the anions are quite similar to each other and in this sense effectively get lumped into the some free energy basin. Similar trends are observed for the larger cations with a low surface-charge density. One interesting aspect that remains to be understood better are how exactly asymmetries in the directionality of the water network in the bulk create similar patterns to those observed in the presence of ions. We caution the reader regarding all these observations since the analysis is conducted with classical point-charge models and only on single ions in solutions.

For future work, it would be interesting to explore how these features change as a function of different salt concentrations with different cation-anion combinations as well as the sensitivity of the results to the use of more accurate electronic structure based simulations.
Indeed, our decision to base the analysis on data obtained from large simulation boxes and long timescales motivated the use of a classical force field. Specifically, we employed the Madrid-2019 force field, which is based on the charge scaling approach and has been shown to reproduce a wide range of structural properties in good agreement with experimental data. However, like many classical models, it falls short in capturing certain experimental observables, particularly those related to dynamics\cite{avula2023}. Achieving agreement between theory/simulations and different experimental properties remains a major challenge in condensed-phase simulations. Recent advances in machine learning force fields and many-body polarizable potentials\cite{avula2023,dissolving_salts_pressure,to_pairnot_to_Scrhan_2024,ML_potential_2024_Bonn,Paesani_halide,ML_potentials_reviewSchran2024,Zhang2023scalable,Paesani_halides_2016} offer promising directions by enabling a more accurate and transferable description of interatomic interactions. The protocol we present here could be applied to analyze trajectories coming from polarizable models or deep neural network–based simulations. In particular, it would be interesting to explore whether incorporating polarization or electronic effects alters the intrinsic dimensionality and whether this, in turn, enables connections to additional experimental observables such as those obtained from vibrational spectroscopy.

Finally, as outlined in the Introduction, our focus in this work is specifically on how ions perturb the local structure of water in their immediate vicinity. One interesting aspect to consider is how non-local structures may play a role in tuning the thermodynamics of ion-solvation. In the case of liquid water under supercooled conditions for example, non-local averaging is needed in order to correctly capture the longer-range spatial correlations\cite{eddy_supercooled_2024}. Other alternative approaches can also be employed to identify spatiotemporal resolutions in an agnostic manner as recently shown \cite{optimal_2024_Pavan_Rcut_deltaT}.

\section{Data Availability}
All materials underlying this study are accessible in the Zenodo repository \url{https://zenodo.org/records/15881029}. The deposit includes (i) the input files for reproducing simulations, (ii) the SOAP descriptor arrays analyzed in the manuscript, and (iii) the Python code used to process the data.

\section{Acknowledgements}
AH thanks the European Commission for funding on the ERC Grant HyBOP 101043272. GS and AH also acknowledge CINECA supercomputing (project NAFAA-HP10B4ZBB2) and MareNostrum5 (project EHPC-EXT-2023E01-029) for computational resources.

\begin{suppinfo}
Analysis of the free energy landscape dependence on the ID given as input of the PAk algorithm for free energy estimation.
Figure showing the normalized distributions of the first ten ion-oxygen distances, calculated along the 1$\mu$s simulations of single ion in water.
Figure showing the normalized distributions of the number of oxygen atoms belonging to the first solvation shell of calcium ion, for the two different Calcium clusters. 
\end{suppinfo}

\bibliography{main}

\newpage
\appendix
\section*{Supplementary Information}

\renewcommand{\thefigure}{S\arabic{figure}}
\setcounter{figure}{0}

\section{Free energy landscape dependence on ID } \label{ID_dependence}

\begin{center}
    \begin{figure}[H]
       \includegraphics[width=\columnwidth]{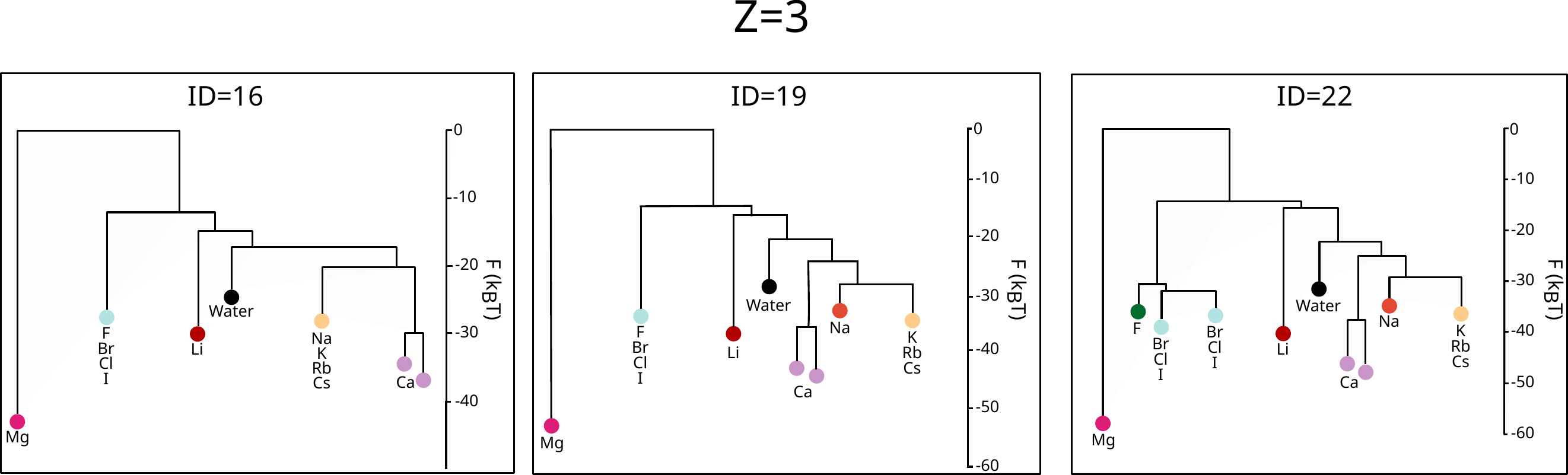}
	       	\caption{Free energy landscapes of the dataset comprising all ionic water environments and bulk water environments, for three values of the intrinsic dimenion given as input of the PAk algorithm for free energy estimation (ID=16,19,22). The $Z$ merging parameter is fixed to three, landscapes are presented using the dendrogram representation.  }
		      \label{ID_dependence}
    \end{figure}
\end{center}

\subsection{Distributions of the distances of the first ten neighboring oxygen atoms  } 
\begin{center}
    \begin{figure}[H]
       \includegraphics[width=\columnwidth]{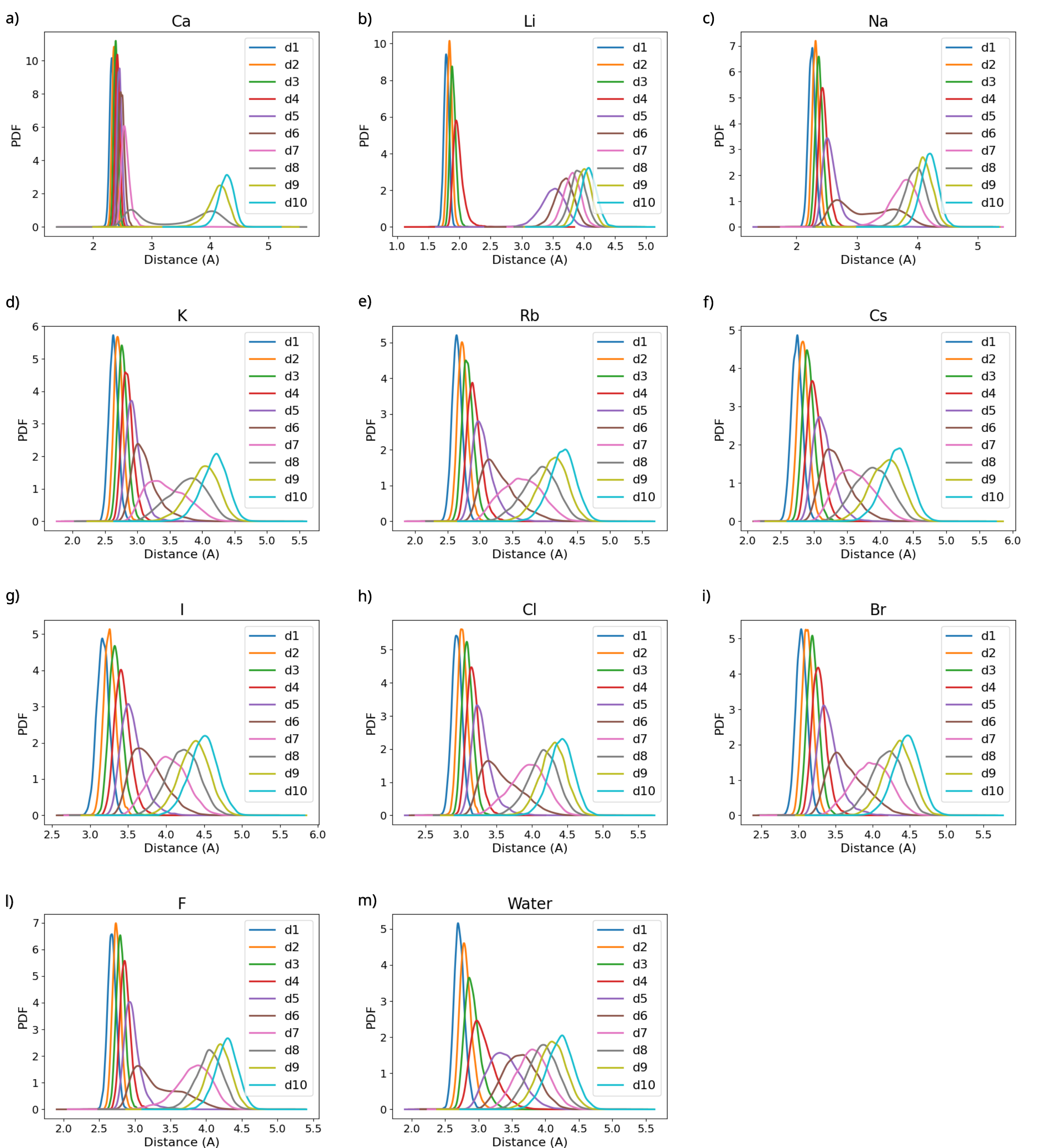}
	       	\caption{Normalized distributions of the first ten ion-oxygen distances, calculated on all the frames of the $1\mu s$ single ion in water simulations.}
		      \label{10dist}
    \end{figure}
\end{center}

\subsection{Subdivision of Calcium frames in two subclusters  } 
\begin{center}
    \begin{figure}[H]
       \includegraphics[width=7cm]{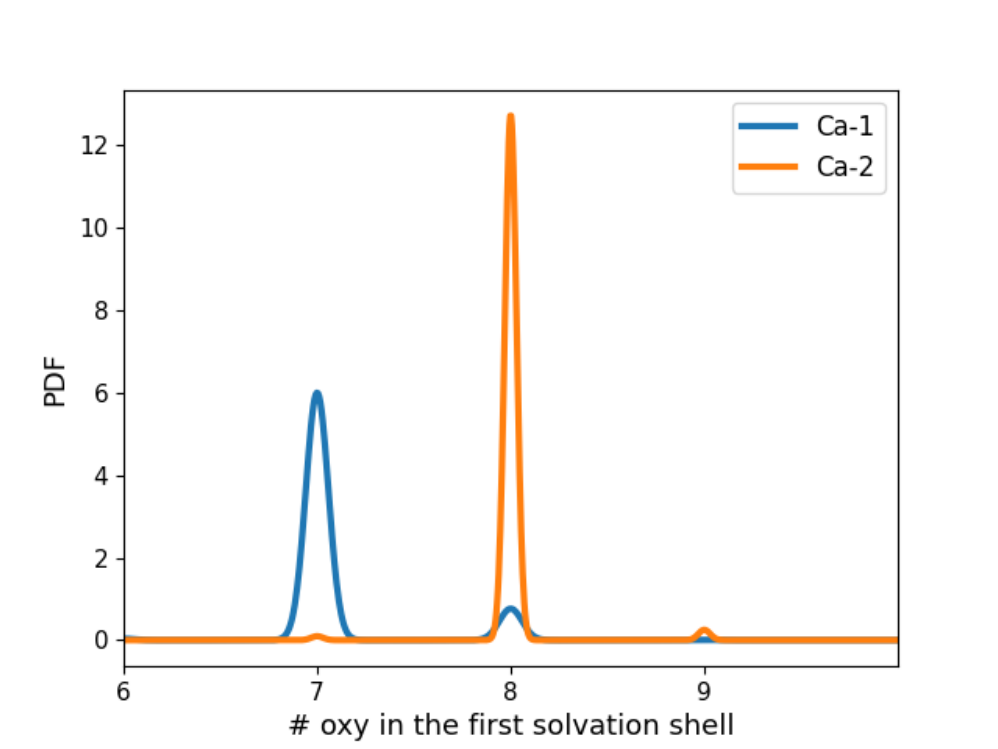}
	       	\caption{Normalized distributions of the number of oxygen atoms belonging to the first solvation shell of calcium ion, respectively calculated on the frames belonging to the two different clusters, both containing water environments around the calcium ion.}
		      \label{Calcium}
    \end{figure}
\end{center}

\end{document}